\documentclass[11pt,twoside,a4paper]{article}
\usepackage{color} 
\usepackage{amsmath}
\usepackage{amsfonts}
\usepackage{amssymb}
\usepackage{graphicx}
\usepackage{fontenc}
\usepackage{times}
\usepackage{mathptmx}

\title{Instructive Review of Novel SFT with 
Non-interacting consituents ``objects'', and attempt  Generalization
to p-adic theory 
}

\author{
Holger Bech Nielsen\\
\\
        Niels Bohr Institute, Copenhagen University\\
Masao Ninomiya\\
        Yukawa Institute for Theoretical Physics , Kyoto University, 
Kyoto 606-0105,
Japan\\ 
and\\ 
Yuji Sugawara Lab., Science and Engineering ,\\
 Department of Physics
Sciences,
Ritsumeikan university
}
\begin{document}
\maketitle
\section{Abstract}
{We have constructed a new formalism for 
describing a situation with {\color{red} 
several (dual) strings} present at a 
time, a {\color{red} string field theory}, 
by means 
of a constituent / a strings from  objects 
picture similar to, but importantly 
different from the ``bits'' by Charles 
Thorn\cite{Thorn}.
Our "objects" 
(essentially the bits) represent rather a 
making a lattice 
in the {\color{red} light 
cone variables} on the string. 
The 
remarkable feature and simplicity of our 
formalism is, that the "objects" do 
{\color{red} NOT interact}, basically 
just run or sit trivially fixed. 
{\color{red} Scattering is a fake} in 
our formalism.
This opens also up for hoping for generalizations 
inspired by hadrons with their partons all having 
 Bjorken variable $x=0$, and thus infinitely many constituents.
The p-adic string is an example.  
}

{\em Corfu Summer Institute 2019 "School and Workshops on Elementary 
Particle Physics and Gravity" (CORFU2019)\\
		31 August - 25 September 2019\\
		Corfu, Greece}

\section{Introduction}
We have worked since long  on a formalism for a string field theory\cite{Kaku3} 
with the possibility of describing an arbitrary number of strings, in 
other words a string field theory, which we called a Novel SFT.\cite{self2}
Our ``Novel SFT'' has same spirit as the before ours developped string from 
bits reformulation of string theory by Charles Thorn \cite{Thorn}. 
In Charles Thorn's theory the string parametrized by the parameter along the 
string usually called $\sigma$ is split into string-bits by discretizing 
the parameter $\sigma$. I.e. each string bit corresponds to a very small 
interval in the parameter $\sigma$. The crucial difference between 
our string 
and that of Charles Thorn 
is that we before going to the splitting into bits consider the wellknown  
conformal gauge choice formalism and use the splitting of the 
fields on the string into right mover and left mover components.
Then the point is that {\bf we} make discretization of the variable 
$\tau +\sigma$ and $\tau -\sigma$ for 
the left and the 
right moving components respectively. 

The real crux of the matter is that we for each of the two ``movers''
only need {\bf one} variable, $\tau +\sigma$ or $\tau-\sigma$ respectively,
whereas say Thorn still has both the discretized variable $\sigma$  and 
the ``time'' variable $\tau$. So while he has to have some 
time development with $\tau$, we basically got rid of the ``time'',
since we only need the variable, which we discretize, i.e. one or the other 
of the two $\tau+\sigma$ or $\tau-\sigma$. In this light it should 
not be so surprising that a main characteristic of our formalism is that the 
time-development for our string-bits, which we since long called 
``objects'' ( which is good to distinguish them from Charles Thorn's 
``string bits'') that they have only a trivial development as time 
goes on. This feature is then a very interesting feature and we want 
to extend and generalize that feature as being of interest in itself.      

We can announce this feature, while thinking of our ``objects'' (analogous 
of the bits)  as consituents of the strings; the strings being then thought 
upon as composed of such ``objects'' by:
\title{\LARGE \bf {Novel SFT with 
Non-interacting consituents ``objects'', p-adic String Generalization 
}}

Here when then also alluded to 
our formalism may bring hopefully some 
understanding of the relation of the usual string theory to the 
p-adic string theories\cite{Branko, padic6}. Well, they should probably then not be called 
string theories, because although the p-adic formalism leads to scattering 
amplitudes analogous to Veneziano models, they are according to the attempt to 
understand them in this article precisely of a different kind of structure 
than the true string: being a long thin structure. Indeed they rather 
share the clumpy structure of the p-adic numbers.  







\begin{figure}
\begin{center}
\includegraphics[width=.6\textwidth]{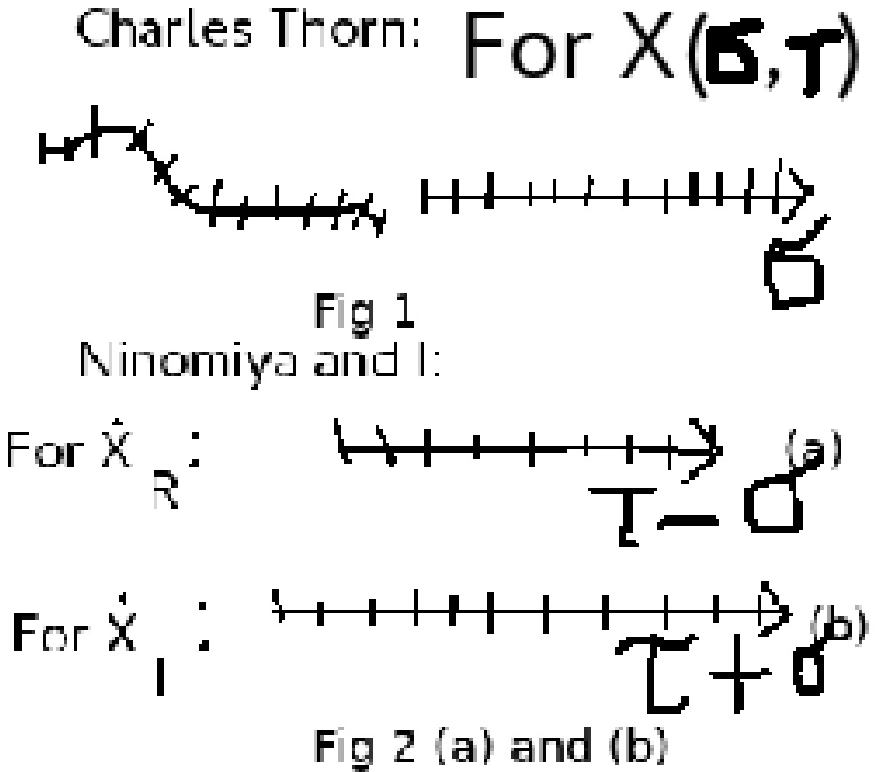}\\
\end{center}
\end{figure}

\subsection{\bf Major Achievement Anti-suggests
putting it into objects:}

The most important gain of (super)string
theory over ordinary quantum field theories
as a theory of everything is:

{\bf {\color{blue} QFT}= Quantum field theories {\color{blue}
ONLY} 
become non-divergent {\color{blue}BY 
RENOMALIZATION};
while the (even made a string field theory)
{\color{red}STRING} theory gives 
{\color{red}FINITE NUMBERS 
FROM THE START} - Veneziano models.}  

{\bf The Wonderful Finiteeness of (Super)String-theory gets Spoiled, 
if Replaced by
a Quantum Field Theory of ``bits'' 
(unless the bits are not normal 
particles)}

\begin{itemize}
\item{\color{blue} Thorn} In Thorn's
the bit (= constituents) do interact, 
but in quite {\color{red}different way 
from ordinary}
quantum field theory particles.
\item{\color{blue} Nielsen-Ninomiya :}
In ours it turns out that the {\color{red}``objects''}
(which is the name we use for our bits) 
{\color{red} \underline {do NOT interact}}. So the 
perturbation 
with the usual divergencies does not
come up.   

\end{itemize} 

\subsection{\bf Overview Ignoring Tecnical Details:}


The reader should get in mind that the connection from our ``objects''
to getting the more physically understandable string is of a new 
type and needs explanation: 

At first have in mind that the state of the whole universe - i.e. 
the structure corresponding to the usual second quantized state in e.g.
usual quantum field theory - is a state in which we have ``objects''
each object having degrees of freedom like that of a scalar particle.
(For a string theory with closed strings there shall be two 
types of ``objects'' ``right-mover type'' and ``left-mover-type''). 
These {\bf ``objects''} are the ones that {\bf do not interact}, so we 
basically 
think of two types of free particles in this description of the analogon 
of the second quantized theory.

Now the rather new idea is that one shall now imagine selecting one right-mover 
type and one left mover type of these particles, which we call ``objects''.
Then we can combine such a pair and construct from that a position, which is 
the sum of the positions of the objects in the pair, and a momentum 
which is the sum of the momenta of the two objects in the pair.

This mathematically constructed {\bf sum of the objects in the pair} shall now 
be identified by essentially a {\bf string bit}  like Chreles Thorns ({\bf at one 
moement} of time at first). That is to say 
that corresponding to such a pair we can - sometimes - attache 
an ``infinitesimal'' bit of a string, which then has its connection to the 
``object'' formalism of ours by having the momentum for example equal to the 
sum of the ``objects'' in the pair.

Now we assume that the objects sit in what we call ``cyclically ordered 
chains'', which first of all means that they are organized as sitting very 
closely along a topologically speaking circle. That means you can imagine 
going arround such a circle / a ``cyclically ordered chain'' visitng one 
object after 
the other one and finally return to the staring one again. 

This means that these ``objects'' sit in a one-dimesnional 
structure, the ``cyclically ordered chain''. 

But now when you attache the string bits - the small pieces of string - 
to {\bf pairs} of ``objects'' - then one easily sees that the pairs will 
correspondingly form a {\bf two}-dimensional sheet.
This two-dimensional sheet is now meant in our picutre to be identified 
with the {\bf time track} of the string(s). That is to say that when we combine
to make string bits all the pairs we can form from consisting of one 
``object'' in the right-mover- cyclically ordered chain and one from the 
left mover one gets not only all the bits arround a closed string at one 
moment of time (think of $\tau$  or $t$ as you like) but rather one gets 
the shape and momentum density say of the string at all different 
moments of time, too.  That is to say that by having the two cyclically ordered 
chains carrying the information for a closed string we have not only the 
information of the string at one moment but get by combining into 
pairs of ``objects'' in different combinations also the behavior of the string 
as time develops!

This is very remarkable since it means that the whole {\bf time-development} 
of the string is {\bf already} basically {\bf built into} the formalism 
in terms of {\bf the objects}. One could say this with the words that 
we have solved the string time development by writting it into 
the ``objects''. But that would be giving the ``object'' formulation
too much honor, because it is wellknow since very long that 
the string dynamics is solvable by writting it into right and left mover 
$X^{\mu}_R(\sigma,\tau)$, and $X^{\mu}_L(\sigma,\tau)$ (see (\ref{rlm})). 
However, we think that this {\bf solving} string theory, is more new 
and more is achieved by our ``object'' formulation, when 
you apply this formulation to a situation with several 
cyclically ordered chains and also several strings resulting from combining 
objects in pairs. 

The truly remarkable observation is that {\bf even when strings 
scatter on each other as they do in string theory, you can claim that 
what goes on in terms of the objects is basically nothing.} the objects  do 
not interact even in the (string-)scattering situation!

\begin{figure}
\begin{center}
\includegraphics[width=0.6\textwidth]{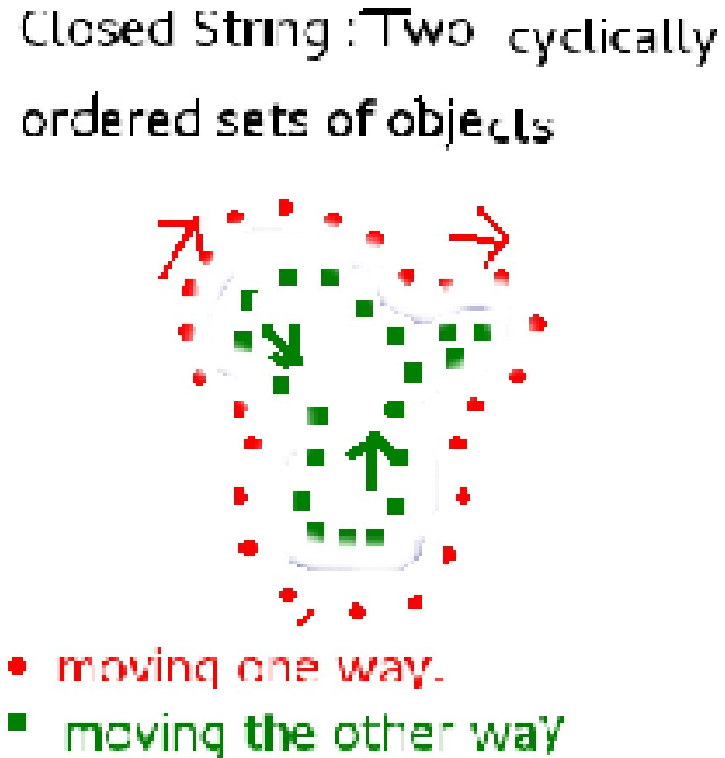}\\
Fig. 3.
\end{center}
\end{figure}

{\bf Notice { two} objects to Each bit 
of String in Charles Thorn Sense}

This ``TWO OBJECTS at each point of the 
string'' is to correspond to the 
wellknown formula for single string 
dynamics in the conformal gauge:

\begin{eqnarray}
X^{\mu}(\sigma,\tau) &=& 
X_R^{\mu}(\tau -\sigma) + 
X^{\mu}_L(\tau + \sigma).\label{rlm}
\end{eqnarray} 
This means that the position and momentum 
of the string-bit sitting at a pair 
of objects  - a rightmover and a 
left mover object - is given as the 
{\color{red} sum} of the two object 
momenta and positions.

This means that the objects, if they 
are considered to be anywhere at all,
are at quite different places from the 
string itself, or the points they 
correspond to. Completely anti-intuitive
for constituents.  
{\bf Think of Right- and Left mover
as distinct degrees of freedom...?}
\begin{figure}
\begin{center}
\includegraphics{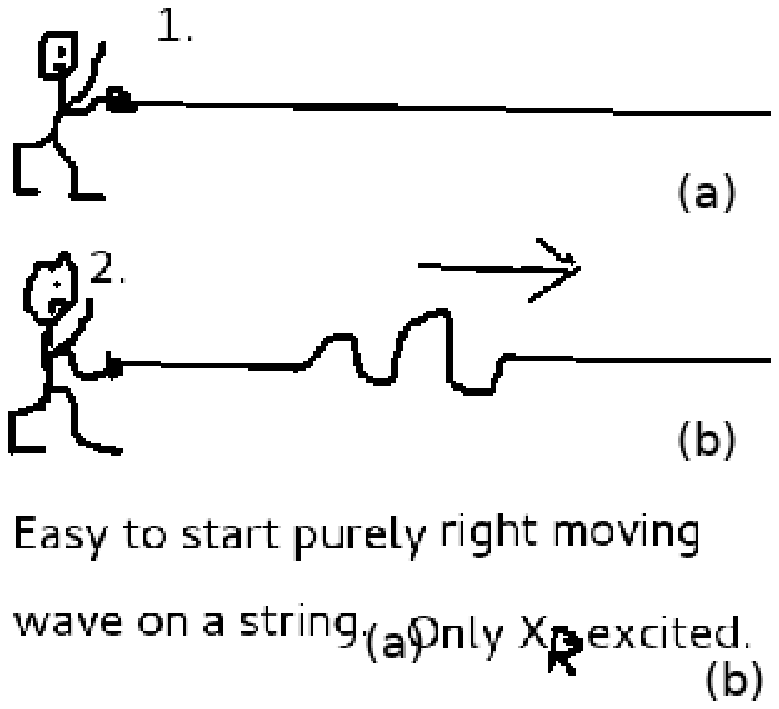}
\end{center}
\hspace{7 cm} Fig. 4.
\end{figure} 

{\bf A little Problem that say $X_R(\tau 
-\sigma)$ not periodic in 
$\tau-\sigma$ but only in $\sigma$.}

\begin{figure}
\begin{center}
\includegraphics{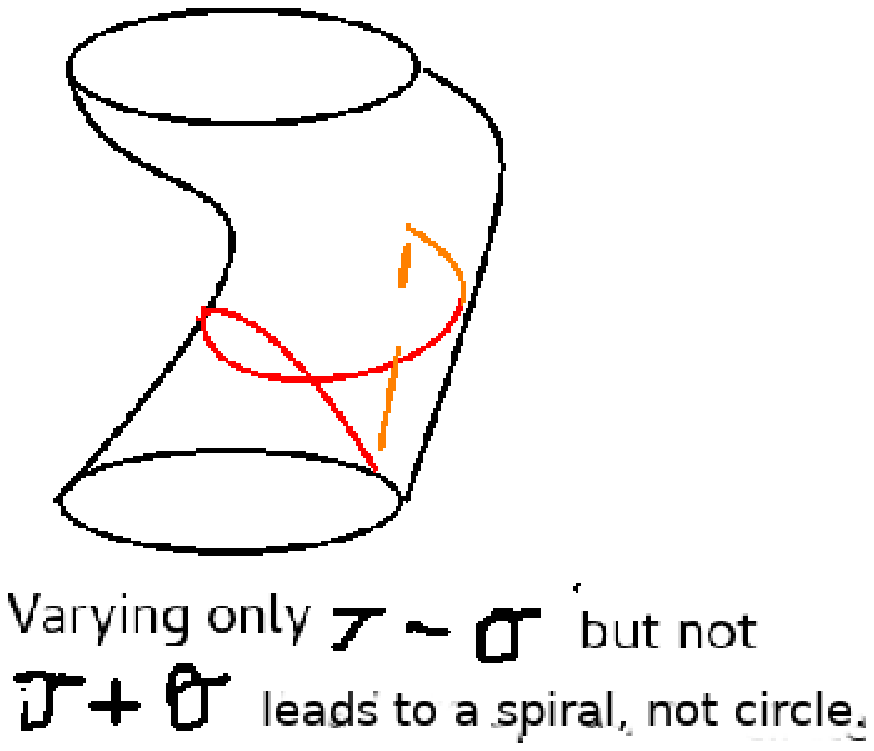}\\
Fig. 5
\end{center}
\end{figure}

{\bf Technical problem, because $X^{\mu}_R$
and $X^{\mu}_L$ are not periodic:}

We have to either:
\begin{itemize}
\item {\color{blue}Declair} that looking 
at these 
right mover and left mover variables 
$X^{\mu}_L$ and $X^{\mu}_R$ 
for the ``objects'' as {\em global}
functions of $\tau +\sigma$ and 
$\tau -\sigma$ respectively is supposed 
to lead to ambiguities, so that we only 
care for local functions in the sense of 
only letting them be defined on smaller 
intervals. {\color{red} only piecewise
functions.}
\item {\color{blue}Or} we identify only the derivatives
or difference of $X^{\mu}_L$ and $X^{\mu}_R$, i.e.
\begin{eqnarray}
J^{\mu}_R(I)&=& X_R^{\mu}(``\tau-\sigma''(I)  
+\Delta/2)- X_R^{\mu}(``\tau-\sigma''(I)
-\Delta/2)\nonumber\\
J^{\mu}_L(I)&=& X_L^{\mu}(``\tau-\sigma''(I)
+\Delta/2)- X_L^{\mu}(``\tau-\sigma''(I)
-\Delta/2)\nonumber\\
\hbox{Here } I&=& ..., -2, -1,
0, 1,2....\nonumber\\
\hbox{ and } \Delta &=& 
``object''distance. 
\nonumber   
\end{eqnarray}
\end{itemize}

More detailed: The $\Delta$ is the ``lattice constant'' 
for the variables $\tau-\sigma$ or $\tau + \sigma$, so that 
we have an ``object'' for each step of length $\Delta$ in these 
variables. Then 
\begin{eqnarray}
J^{\mu}_R(I) \approx \Delta*\dot{X}_R^{\mu}(``\tau-\sigma''(I))\nonumber\\
J^{\mu}_L(I) \approx \Delta*\dot{X}_L^{\mu}(``\tau-\sigma''(I))\nonumber.
\end{eqnarray}

{\bf We then want to Arrange:}

 {\bf Each ``object''
a Full Physical System (a set of variables
and their canonically conjugate)}

But have in mind:

The right and left mover variables 
$\dot{X}^{\mu}_L$ and $\dot{X}^{\mu}_R$  
have in  
string theory  a derivative of delta 
function commutation rule with themselves:
\begin{eqnarray}
\left [{ \dot{X}_L^{\mu}(\tau_{L1}),
\dot{X}_{L}^{\nu}(\tau_{L2})} \right ] &=&
-i g^{\mu\nu}\delta'(\tau_{L1}-\tau_{L2})
\nonumber \\
\left [{ \dot{X}_R^{\mu}(\tau_{R1}),
\dot{X}_{R}^{\nu}(\tau_{R2})} \right ] &=&
-i g^{\mu\nu}\delta'(\tau_{R1}-\tau_{R2})
\nonumber \\
\hbox{ where }&&\nonumber\\
\tau_R&=& \tau-\sigma \nonumber \\
\tau_L&=& \tau + \sigma \nonumber 
\end{eqnarray}
 
So one object would at first not commute 
with its neighbors. 

{\bf We move the Information on Oddly 
Numbered Objects to Neighboring Evenly 
Numbered Ones to make up Conjugate 
Variable to say $\dot{X}_R(I)$.}

We found a way to achieve the following 
wishes before identifying (half) the 
objects 
with particles in a second quantized 
field theory. Think of at first having 
the objects be described just by $X_R(I)$
and $X_L(I)$, then modify them somehow
to achieve:

\begin{itemize}
\item Those objects, which we end up 
keeping in the formalism, shall if  
associated with one variable also be 
associated with its canonically conjugate
one together with it.
{\color{red} Objects have full sets of 
degrees of freedom.}
\item All variables associated with one
object shall commute with those 
associated with an other object.
{\color{red} Different objects commute.}
\end{itemize}

{\bf Wishes about to construct 
Object-degrees of freedom:}
\begin{itemize}  
\item We select a subset of objects still
so that all information in $\dot{X}_R(\tau_R) $
is modulo the discretization in the 
``kept'' ``objects''. {\color{red} The 
kept right objects carry all the 
right mover information; and the left 
carry the left mover information.}   

\end{itemize}


{\bf The Commutator of the $\dot{X}_R(I)$
of the $\delta^{'}$ form.}

\begin{figure}[h]
\begin{center}
\includegraphics {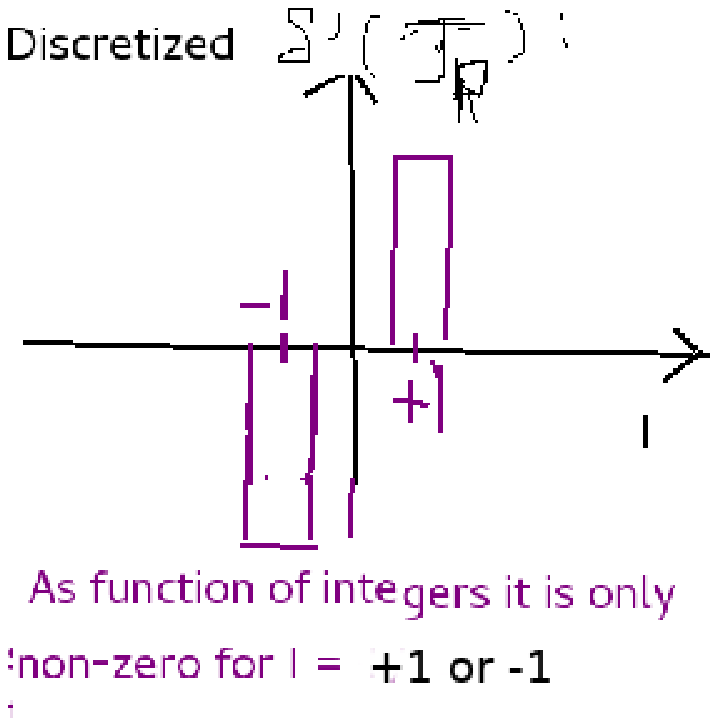}\\
Fig. 6
\end{center}
\end{figure}

{\bf Because the $\delta'(\tau_R)$ 
discretizes to a function only non-zero 
in two neightboring points to zero, we 
could achieve full commutation by 
leaving out all objects with an odd 
number.}

So we shall seek to put in all 
information from $X_R$ or from $\dot{X}_R$ 
in to {\color{red} only the even numbered
objects.} But the odd ones are 
essentially the conjugate 
of 
the even ones. In fact we can and shall
choose the odd objects to be written 
as differences of the conjugate variables
for the even neighbors. 

\begin{figure}
\begin{center}
\includegraphics[width = 0.6\textwidth]{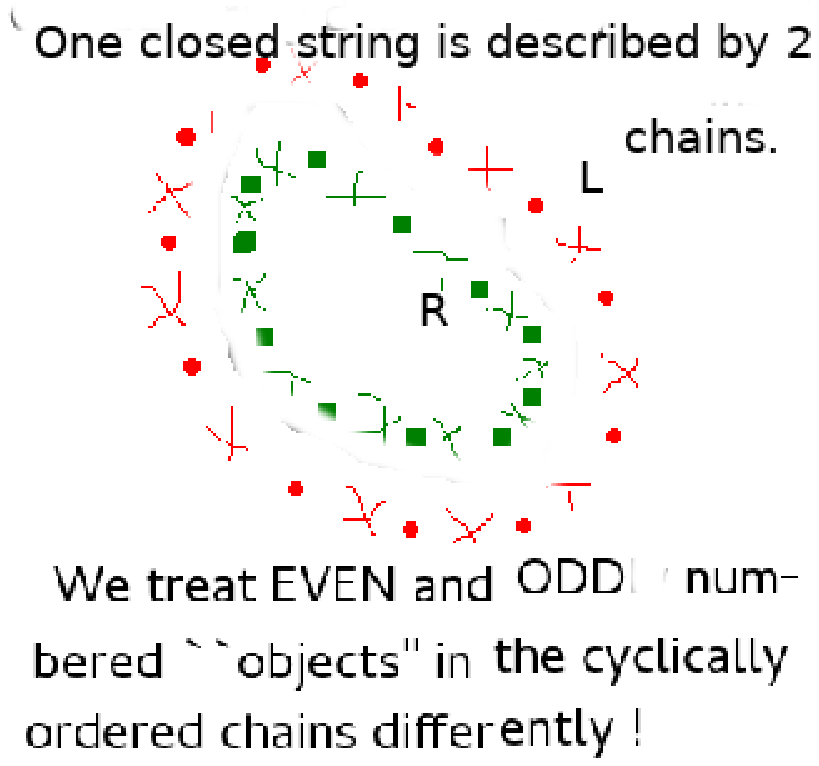}\\
Fig. 7
\end{center}
\end{figure}

{\bf We achieve the wishes by the 
following ansatz:}

We define for each {\em even numbered}
($I$ even) 
object a variable set $J_R^{\mu}(I)$ and 
the canonically conjugate set $\Pi_R^{\mu}
(I)$ (here leaving out a quite analogous 
left set with $L$ instead of $R$). We call
$\Delta$ 
the discretization step:
\begin{eqnarray}
J_R^{\mu}(I)&=&
\dot{X}_R^{\mu}(I)\Delta=\nonumber \\  
=X_R^{\mu}(\tau_R(I+1/2) -
X_R^{\mu}(\tau_R(I-1/2)&& \hbox{  (for
$I$ even)}; \nonumber \\
-\pi\Delta *(\Pi_R^{\mu}(I+1)-
\Pi_R^{\mu}(I-1))& =&
\dot{X}_R^{\mu}(I)\Delta=
\nonumber  \\
= X_R^{\mu}(\tau_R(I+1/2) -
X_R^{\mu}(\tau_R(I-1/2)&& \hbox{  (for
$I$ odd)} \nonumber \\
\end{eqnarray}

Note: We only use $\Pi_R^{\mu}(I)$ and 
$J_R^{\mu}(I)$ here for {\em EVEN} $I$.

{\bf We got a way to put the information
of a free closed string solution of 
equations 
of motion into that two sets of 
 (infinite) numbers of ``even objects''
with their $(J_R^{\mu}(I), \Pi_R^{\mu}(I))$
$(I= ...,-4,-2,0,2,4,...)$.}

Summary: We can describe {\bf one} closed 
string by two sets, $R$ and $L$, of 
cyclically ordered ``even numbered 
objects''.

Each ``object'' 
has in the 
bosonic string 
theory one degree of freedom - a 
$J_R^{\mu}(I)$ and the canonically 
conjugate $\Pi_R^{\mu}(I)$ one  - 
for each of the 25+1 dimensions.
(modulo some troubles with gauge choosing
...)   

{\bf Main Point: We  Second Quantize the 
Objects; and There can be Many Strings 
in the Same Quantum Field Theory}

\begin{itemize}
\item First describing {\bf one} string
you let it correspond to a {\color{red}quantum field 
theory} in which you have the two 
cyclically ordered chains of ``objects''
(= the particles in the quantum field 
theory (with massless non-interacting 
spin zero particles)).
\item This gives the obvious idea to 
describe {\color{red} several strings},
namely by just putting more couples 
of cyclically ordered chains up in the 
same quantum field theory Hilbert 
space(s). I.e. you got a {\color{red}
``string 
field theory''} (= second quantized 
string theory like Kaku and Kikkawa\cite{Kaku3} 
or Witten\cite{Kaku3}...) {\color{red} for free!}
\end{itemize}
\subsection{ Scattering of Strings is a Fake.}

We managed - though with some 
technical difficulties - to obtain
string scattering as a {\color{red} 
fake}, meaning that the objects 
themselves do {\bf not} do 
anything during the scattering, but 
we/the physicists {\color{red} reclassify
the objects into {\bf new cyclically
ordered chains}}, and then we have 
a formal scattering, although in the 
physics of our model for non-interacting
``objects'' nothing can happen.     

{\bf In String Theories with Open Strings
Only One type of ``Objects''; while 
only closed ones have $R$ and $L$.}

\begin{figure}[h]
\begin{center}
\includegraphics{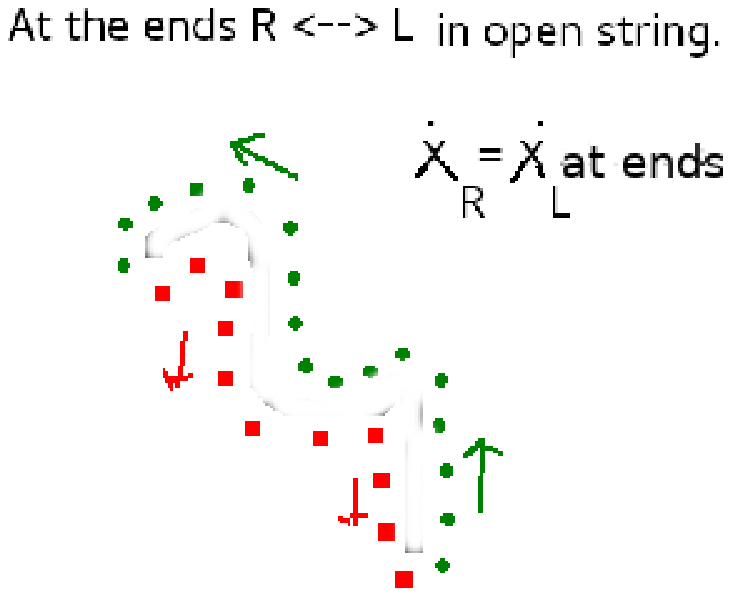}\\
One open string\\
Fig. 8\\
\label{Fig8}
\end{center}
\end{figure}

{\bf The Constraints $\dot{X}^2 +(X')^2 
\approx 0$ and $\dot{X} \cdot X' \approx
0$ look nice in ``objects''.}

The usual constraints that
  $\dot{X}^2 +(X')^2$ 
 and $\dot{X} \cdot X'$ 
should 
be ``weakly''  $0$ take a nice form 
in the variables used for our ``objects'':
\begin{eqnarray}
\dot{X}_R^2 &=& \left ( \dot{X}_R^{\mu}
\right )^2 \approx 0 \nonumber \\
\dot{X}_L^2 &=& \left ( \dot{X}_L^{\mu}
\right )^2 \approx 0
\end{eqnarray}
Here we used $\approx$ to denote the 
``weak'' equality.

\begin{figure}[h]
\begin{center}
\includegraphics[width=0.6\textwidth]{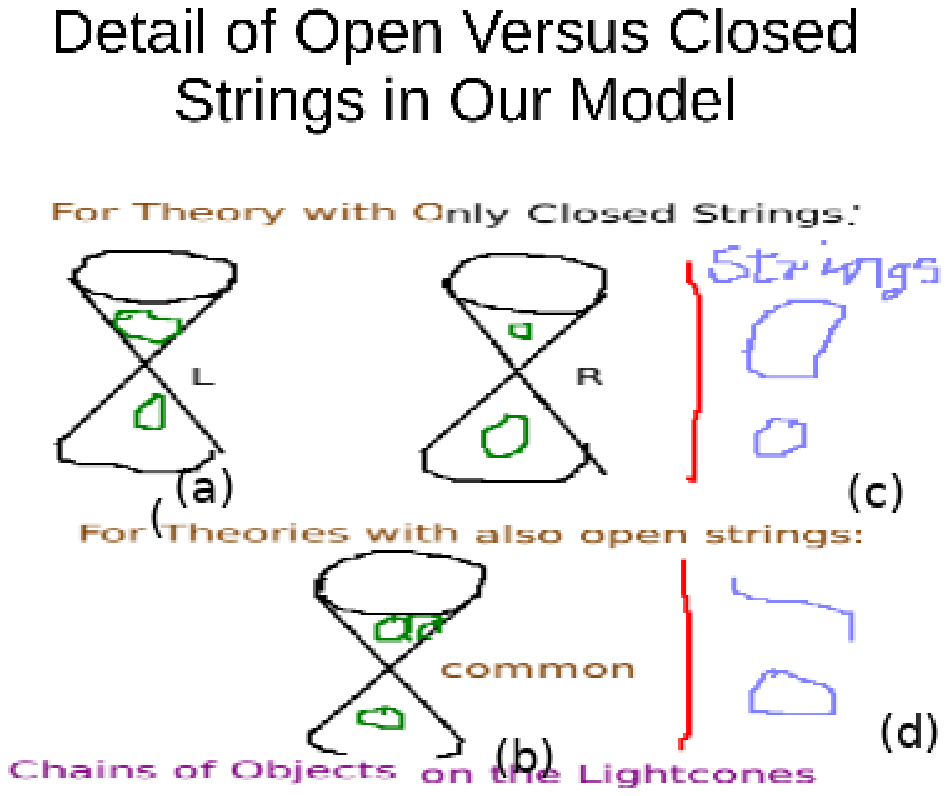}\\
String in terms of objects:\\
 closed (a),\\
open and closed (b)\\
The string of $X_{\mu}$ in 25+1 dimensions: \\
closed (c),\\
 and open and closed (d) respectively.\\
Fig 9 
\end{center}
\end{figure}


\subsubsection{\bf {\color{red} Stringiness 
(one-dimensionaly touching of objects) is
put in via the 
States of the 
``Objects'' 
}}

Since we have no interaction between the ``objects''
such an interaction cannot be used to keep or make  the objects 
sitting in a chaine. Since they only move trivially though 
they will continue to ``sit''( or better run) as forming a 
continuous chain, provided they have started forming such 
a chain. In this sense the only way to assume the objects to 
form continuous chains would be to just assume it as an inittial 
condition, being true - say in the beginning - and thus continuing 
being true forever (from the trivial development).

\begin{figure}[h]
\begin{center}
\includegraphics[width=0.6 \textwidth]{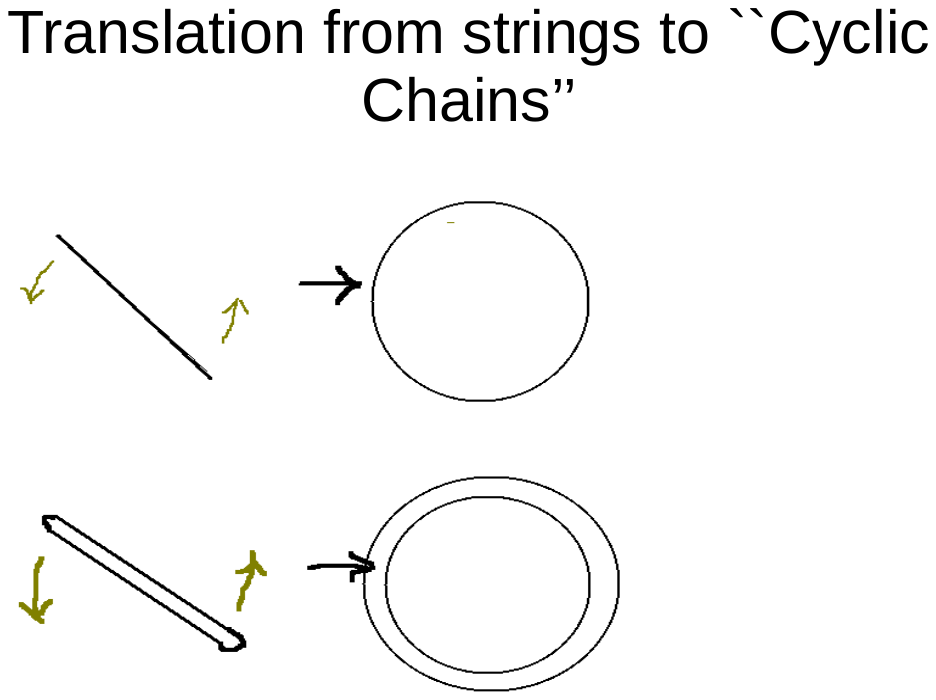}\\
Fig. 10
\end{center}
\end{figure}

Our ``String Field Theory'' model
is basically the second quantized theory
described as particles (best in 
a timeless universe), and then with the 
constraints $\dot{X}^2 
+ (X')^2 \approx 0$ and $\dot{X}\cdot
X'\approx 0$ imposed.

In that model, a quantum field theory 
of just non-interacting massless 
particles, there is no stringiness at 
all! {\color{red} It is only by the 
(initial) states of the objects 
{\bf considered} that there is any 
allusion to the hanging together 
to one-dimensional chains.} Something
that at the end leads to the stringy 
hanging together.

{\color{blue} The string is in the 
initial - and also final conditions 
we need - conditions only!} 

(When we derive our Veneziano model amplitudes 
from our Novel SFT\cite{self2}, then it is crucial for reaching  the 
Veneziano model, that also the ansatz states for the strings 
in the final state - the out going strings - are chains with 
the same Gaussian wave function and associated continuity 
behavior having at least some stringiness in it. Thus we assume 
basically  something about the state of the future, namely 
a certain stringiness.)
   
\subsubsection{\bf A single Open String is described by
just {\color{red} one} Cyclically ordered chain of
``even objects'':}
\begin{figure}[h!]
\begin{center}
\includegraphics[scale=0.4]{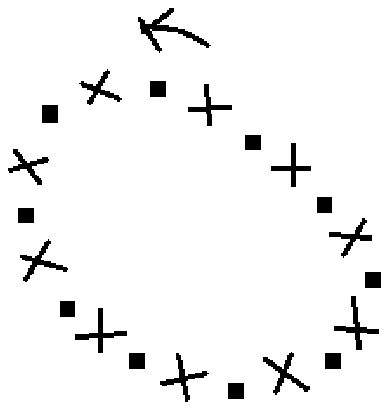}\\
(a)
\end{center}
\begin{center}
\includegraphics[scale=0.4]{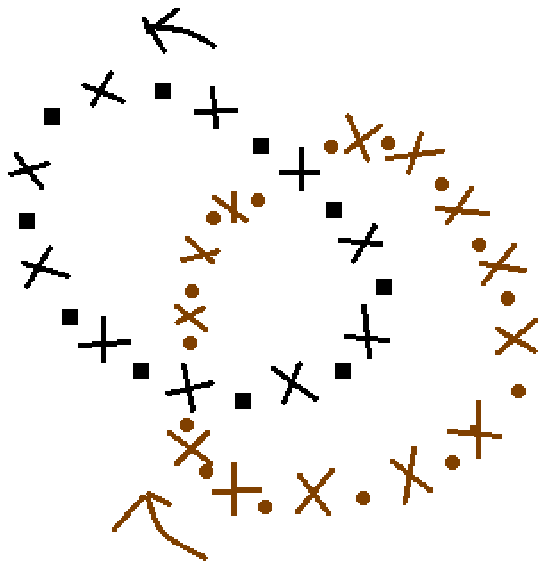}\\
(b)\\
Fig 11-(a) and (b)\\
On (b) cyclically ordered chains of objects for two open strings. 
\end{center}
\end{figure}
\begin{figure}[h!]
\begin{center}
\includegraphics[scale=0.4]{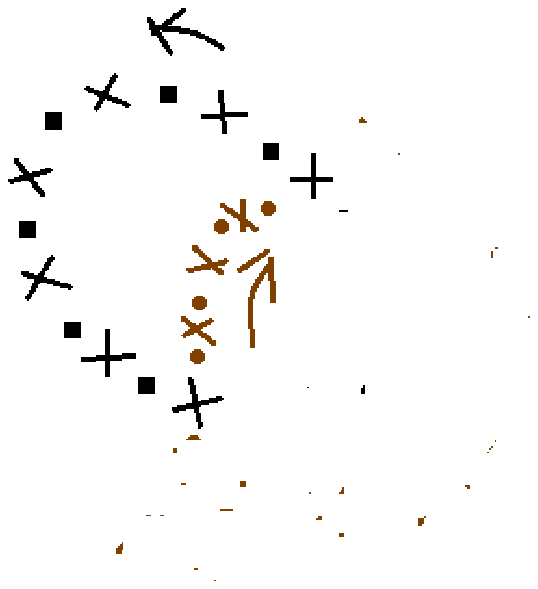}\\
(a)
\end{center}



\begin{center}
\includegraphics[scale=0.4]{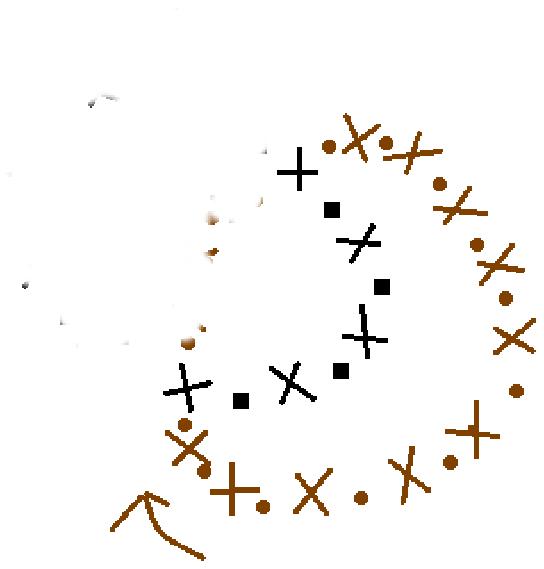}\\
(b)\\
Fig 12 (a) and (b)
\end{center}
\end{figure}

{\bf Now Add One more Open String 
described also as a Cyclically Oredered 
Chain of Objects.}
 
{\bf If the two cyclically ordered chains cross in two places, we can 
interprete it
as corresponding to two open strings in
different ways.} Indeed wecan first interpret it as a couple 
of cyclically ordered chains in Minkowsky space in two ways:
Indeed we symbolicaly show in fig 11 (b) two cyclically ordered chains,
a black and a brown say, having two points in common. By using just the 
points, objects, one can see that we extract from the combination of these 
objects the two combinations of objects shown on Fig 
12 (a) and 
fig 
12 (b) can be extracted. These cyclically ordered chains 
fig 
12(a) and 
12 (b) are each of them along some piece at least 
different from the two illustrated in fig 11 (b). Thus one can 
out of two in two points crossing cyclic chains extract quite differnt ones.
But now there correspond to different cyclically ordered chains open strings 
with different shape and time development.

{\bf The Rest of the Objects also 
form a Cyclically ordered chain menaning
it gives a String.}

{\bf Without Anything happening except in 
Phantasy two strings scattered to two other strings!}

\begin{figure}[h]
\begin{center}
\includegraphics{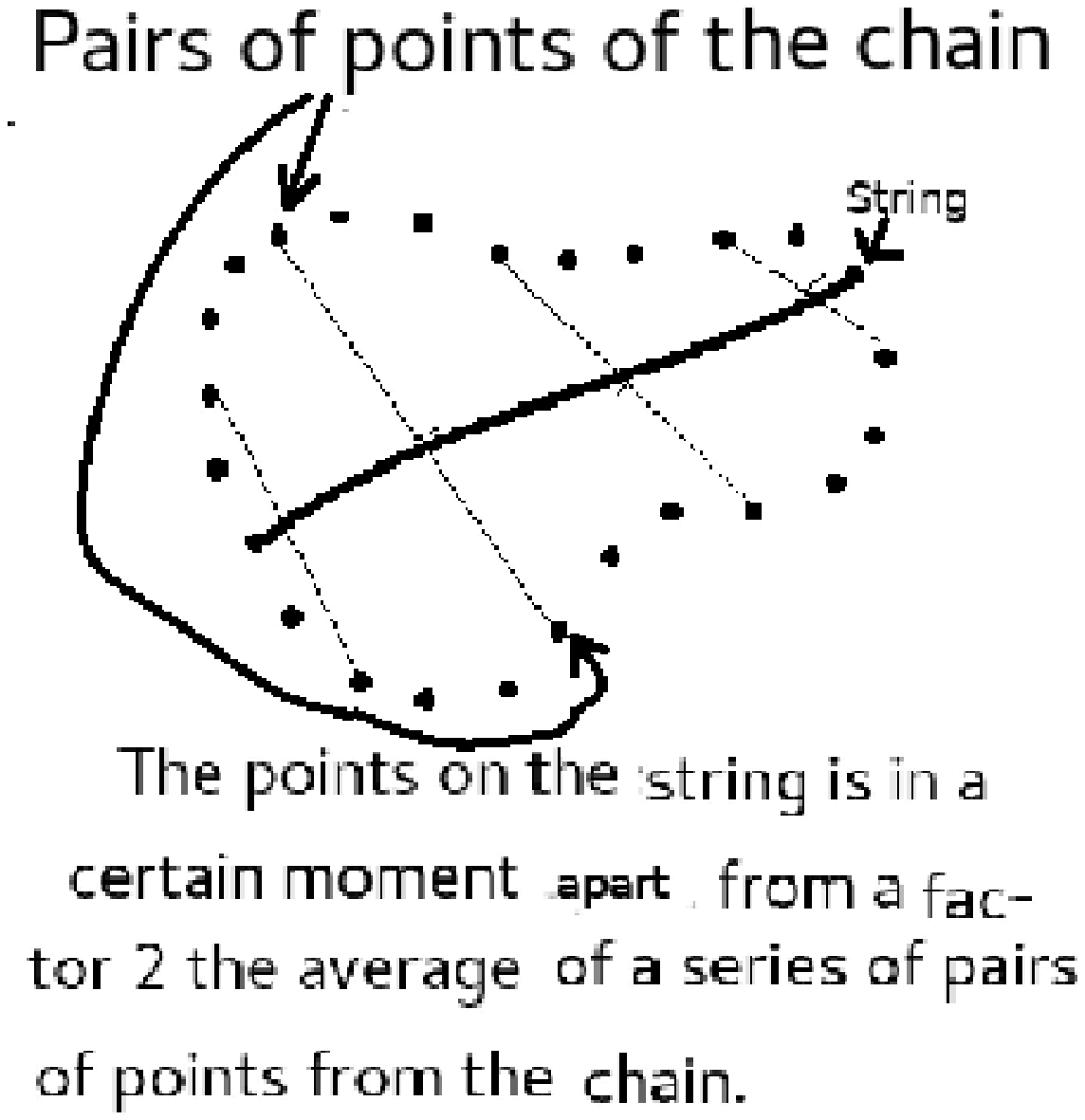}\\
   
Fig. 13
\end{center}
\end{figure}
The reader may keep in mind the way one obtains the open string from a 
cylcically  ordered chain of objects. In fact it is so that sum the 
space-time positions of any pair of the objects to get that of 
a space-time point for a point on the (open) string. 
Whether we sum such a pair of point-coordinates or we take the 
average of theem is of course only a convention of a factor 2.
So you could then say the events through which the the string passes 
is the collection of all the averages of pairs of points on the 
cyclically ordered chain. Thinking of the situation in a moment of time
we can claim as drawn in the figure to give at least the crude idea
%
that the string is gotten by  aseries of pairs of 
objects - paired with each other by a series of locally approximate 
parallel lines. The string goes through the midle points of 
all the pairing pieces of lines.

If one has couple of double crossing cyclically ordered chains,
one can expect that in some Lorentz frame you can find the 
piece of line connecting the two crossing points and on the 
midle of that connection line piece will be the 
crossing point of the open strings for the appropriate moment:
\begin{figure}
\begin{center}
\includegraphics{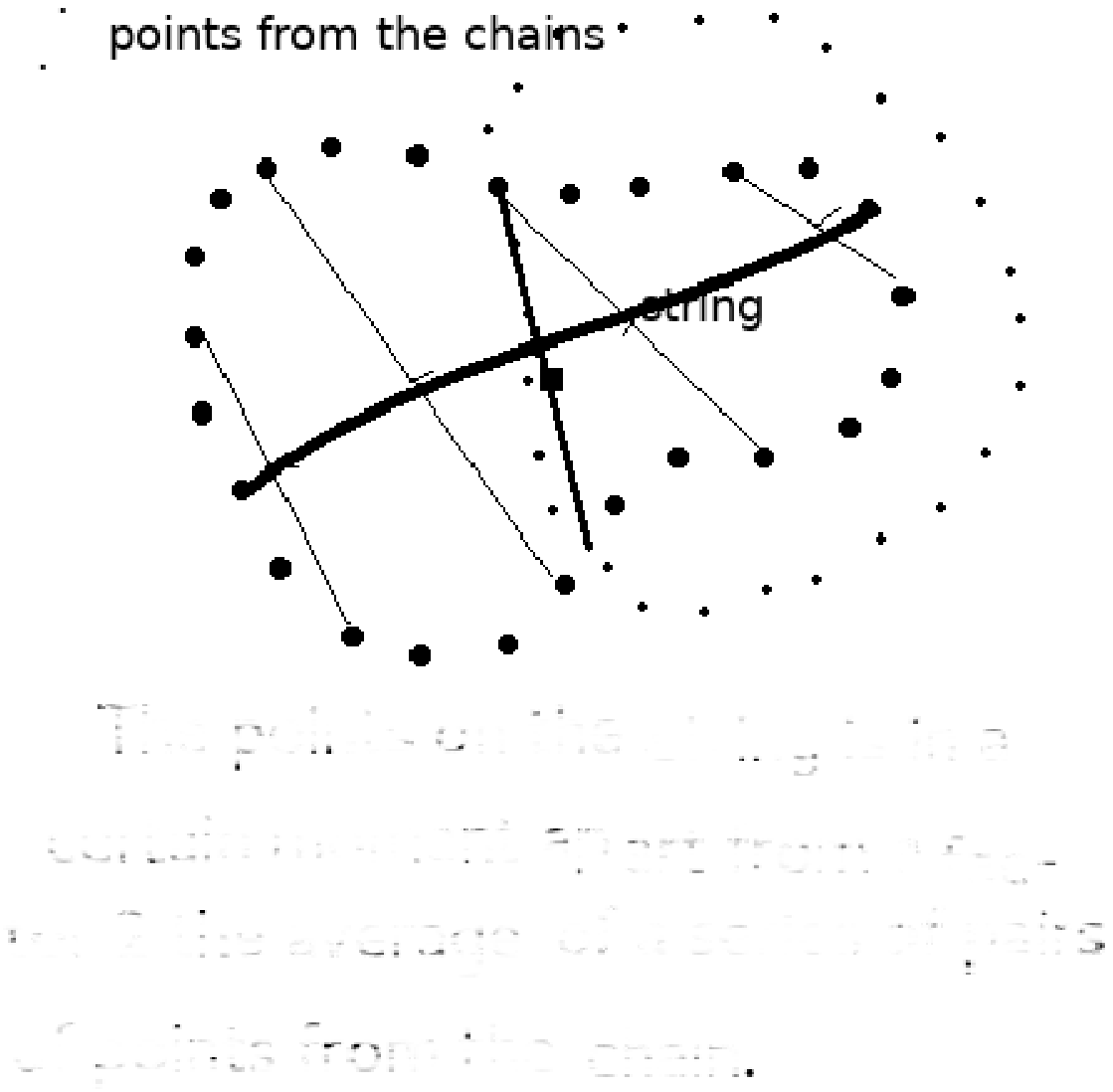}\\
Fig 14.
\end{center}
\end{figure}
The idea of ``faked scattering'' now
is that we split the combined sets of the 
two cyclically ordered chains into a 
couple of cyclically ordered chains 
{\color{red} in a new way!} We take 
part of the chain corresponding to the 
first of the two initial strings and 
combine it with a part of the one that 
corresponds to the second string.
\section{Achievements}
\subsection{ In principle we rewrote String Theory,
but still would like to check if it is true.}

To argue that our 
string theory really
is correct, one would first see that essentially
from locality one can see that indeed the system
of ``objects'' will {\bf not} change in 
a true scattering of dual strings.

{\color{blue} Think say of a couple of strings 
scattering classically by ``exchange of tails''.
Then the image of the $\dot{X}_R^{\mu}$ will only 
change on a nullset.}
{\bf By locality one argued the image 
of the $\dot{X}_{R/L}^{\mu}$ remain the same
( mod. nullset)}
\begin{figure}
\begin{center} 
\includegraphics{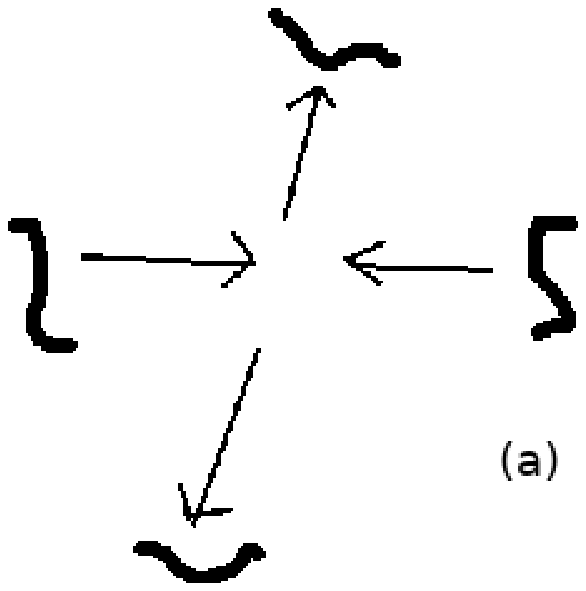}\\
\end{center}

\begin{center}
\includegraphics{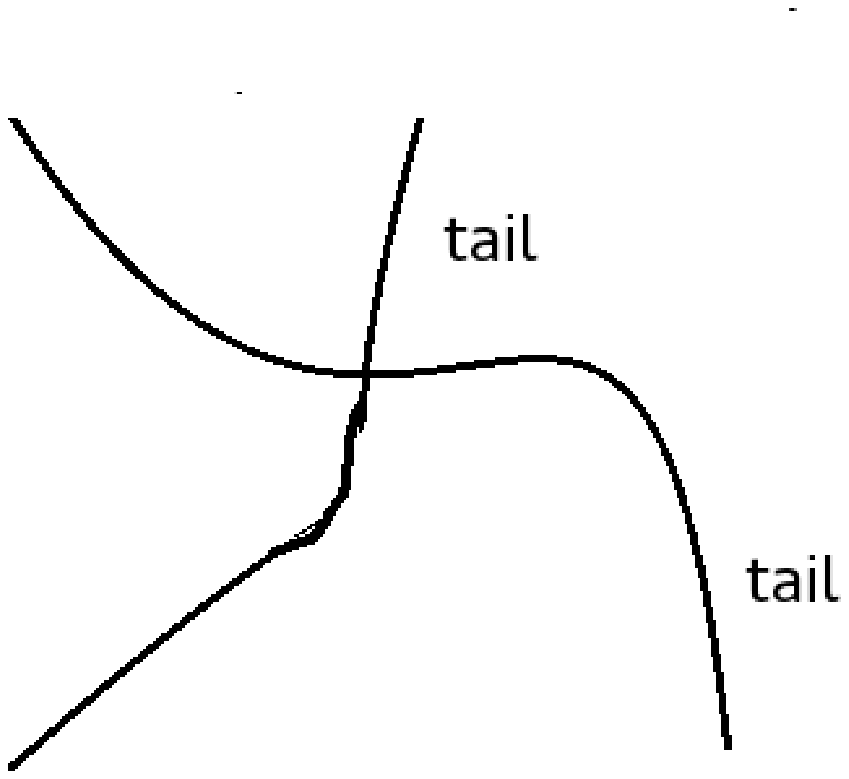}\\
Two open strings hit in a point, and exchange tails.\\ 
Fig 15 (a) and (b)\\
\end{center}
\end{figure}
{\bf In fact we calculated:}
\begin{itemize}
\item Spectrum of a string using our Novel String Field 
Theory Model.\cite{self2, self8, NN7}
\item The (``faked'') scattering amplitude,
expecting to get the Veneziano model; but 
got only one out of three terms in the infinite
momentum frame gauge choise. 
\item Including the possibility of negative 
energies for the constituents(objects) 
- like in Nambu-Bethe-Salpeter\cite{BS} equation - 
we at least glimps a way to get all three terms of the Veneziano amplitude. 
 
\end{itemize}
\subsection{ The Veneziano Model Derivation Quickly 
comes to Technical String Calculation\cite{self2}:}

In spite of the fact that our model being only a 
quantum field theory formalism having no 
stringiness proper in itself, but only in
initial and final state input, the derivation 
of the Veneziano model from our model quickly 
goes into the track of string theory:
\begin{itemize}
\item We begin the calculation by representing 
the external string by wave functions in terms 
of ``objects'' being derived from a functional 
integral through an imaginar time. 

\item Next we must take the overlap between the inital
state - of some strings giving some cyclically ordered 
chains - and the final state. 

\item This leads to gluing 
together the space-time or rather $(\tau,\sigma)$ regions
from the various imaginary time developments. 
\item The glued together regions for definition 
of the functios over which to functionally 
integrate  leads 
to complex surfaces 
which becomes just 
like the corresponding functional integral regions  in 
usual string theory.
\item The counting of the various ways to 
identify the systems of objects in initial 
and final states leads to precise integration measures 
for the Veneziano model integrals.
\end{itemize}

\subsection{ The ``usual surfaces'' in string theory.}

\begin{figure}
\begin{center}
\includegraphics
{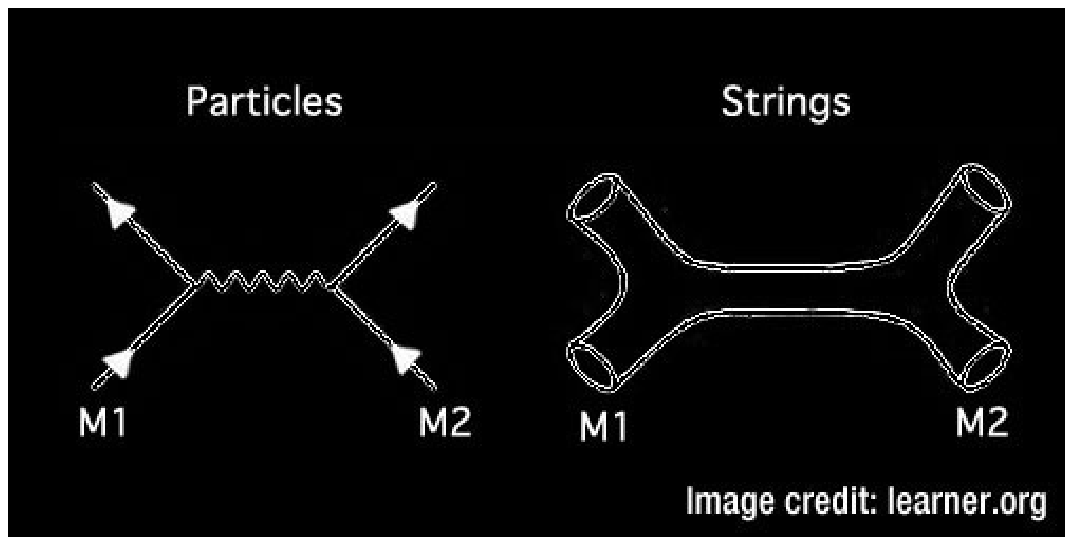}\\
\label{14} 
Fig 
16
\end{center}
\end{figure}

{\bf The basic second quantized Hilbert space for our string field theory
is only particle second quantized space ${\cal H}$.}

On this ordinary free particle  second quantized Hilbert space ${\cal H}$
one can create particles (free and massless) with creation 
operators denoted $a^{\dagger}(J^{\mu})$, where we can think of the 
26-vector $J^{\mu}$ (which is the main variable for one of our objects/bits)
as say the 26-momentum (really we may only need 24 components in infinite 
momentum frame) for the free massless particle that can be created in our 
basic model Fig 2.

{\bf Formulation of Second quantized ``strings''}
\begin{itemize}
\item The string states are created by acting on this space ${\cal H}$ with 
an in the limit infinite number of creation operators $a^{\dagger}(J^{ \mu})$  
associated with  
a series $J^{\mu}$-values (26-vectors) as taken on by 
our type 
``objects'' and weighted with a wave function $\Psi(J^{\mu}_0, J^{\mu}_2,...
, J^{\mu}_{N-4}, J^{\mu}_{N-2})$ for a single string described as consisting 
of a cyclically ordered chain of even numbered ``objects'':
\begin{eqnarray}
&&|string_{\Psi}>=\nonumber \\ 
&=& \int\cdots \int {\cal D}J^{\mu}*\nonumber \\
&&* \Psi(J^{\mu}_0, J^{\mu}_2,...
, J^{\mu}_{N-4}, J^{\mu}_{N-2})*\nonumber \\
&& *a^{\dagger}(J^{\mu}_0)a^{\dagger}(J^{\mu}_2)\cdots
a^{\dagger}(J^{\mu}_{N-4})a^{\dagger}(J^{\mu}_{N-2})|``vac''>.\nonumber 
\end{eqnarray}

The state $|``vac''>\in {\cal H}$ is a priori the empty 
state, meaning  without any of the scalar free particles 
($\sim$ our objects).
\end{itemize}

\subsection{ Really the vacuum $|``vac''>$ should be Rough}

 We ran into troubles of only obtaining one out of 
three Euler Beta-functions for the scattering amplitude by using infinite 
momentum naturally combined with such a no particle vacuum used for 
$|``vac''>$. 
This problem is supposed to be cured by a more complicated vacuum state
(the rough Dirac sea for Boson \cite{DS5}), so that one can both 
add and subtract energy to 
the back ground state $|``vac''>$ used. 
{\bf Crux: We can make many string states trivially:}

A single string creation operator in our scheme in terms of 
object-creation operators $a^{\dagger}(J^{\mu})$ has the form
\begin{eqnarray}
&&{string}_{\, \Psi}=\nonumber \\ 
&=& \int\cdots \int {\cal D}J^{\mu}*\nonumber \\
&&* \Psi(J^{\mu}_0, J^{\mu}_2,...
, J^{\mu}_{N-4}, J^{\mu}_{N-2})*\nonumber \\
&& *a^{\dagger}(J^{\mu}_0)a^{\dagger}(J^{\mu}_2)\cdots
a^{\dagger}(J^{\mu}_{N-4})a^{\dagger}(J^{\mu}_{N-2}).\nonumber
\end{eqnarray}

With that we can simply make a several string state by acting 
with several of these operators on a vacuum 
$|``vac''>$:
\begin{eqnarray}
&&|1,2,...,n>=\\
&=& string_1^{\dagger}string_2^{\dagger}\cdots string_n^{\dagger}|``vac''>
\end{eqnarray}

\subsection{ Got Veneziano model from Fake scattering!}

When we used infinite momentum frame (for gauge choice)
we got, but interestingly enough only quite right for 25+1 
dimensions, (for our bosonic string constituent SFT) for the scattering 
amplitude $A(s,t,u)$:

\begin{eqnarray}
A(s,t,u)&=& C*B(-\alpha(t), -\alpha(u))\nonumber \\
\hbox{where e. g. }\alpha(t) &=& \alpha(0) + \alpha'*t\nonumber \\
\hbox{and } \alpha(0) &=& -(-1/\alpha') = 1/\alpha' \nonumber 
\end{eqnarray}
We did not determine the overall coefficient $C$ in the mentioned calculation.

\section{Perspective}

{\bf Perspective: String theory may be considered as a mathematical 
construction on a free 
quantum field theory (for  massless spin zero particles)}
 {\bf Remarkable} with our  
 non-interacting constituents (=objects= our string bits)  
string field theory model: 
\begin{itemize}
\item The basic SFT Hilbert space is just that of free massless 
{\bf particles} and carries no signal of strings a priori in it.
Rather {\bf the strings come in only via the 
objects being 
inserted in} (quantum fluctuating){\bf chains}!
\item {\bf Nothing happens to the objects} under the scattering!
It is {\bf only}, that one has a {\bf non-zero overlap between the object 
states 
corresponding to the strings in the ``incoming'' state of 
strings and in the ``outgoing'' one!}  Rather as if the strings 
are just a  clever($\sim$ {\bf artificial}) construction in a fundamental 
world, most naturally 
 interpreted as a particle theory. 
\end{itemize}

\section{Genralizations}
{\bf Generalization of non-interacting constituent idea}

Let us suppose we {\bf seek a theory that is finite} by 
looking e.g. at hadrons with their {\bf rapidly falling off}
amplitude for {\bf large} transverse {\bf momenta}.

{\bf If} they really {\bf fell off} exponentially they would in loops 
give rise to {\bf convergent loop} corrections and there would be 
{\bf no} ultraviolet {\bf divergence propblem}; {\bf but } we now 
know that 
the hadrons contain {\bf partons} and because of that 
emit {\bf 
less strongly falling off} particles (hadrons). 

{\bf \color{red} But if all the partons had Bjorken variable\cite{Bjorken6}
 $x=0$, then 
there would be no energy for the partons to scatter on each other 
and the exponential fall off would be preserved!} 

{\bf How the soft/exponential cut off of the Veneziano model comes in say 
$p_{transverse}$ 
 in our formulation. }

{\bf  Object-picture from the wave function of the string in terms of
consituents.}

Since the main thing - to get soft cut off to {\bf avoid divergences} 
hopefully - comes from the wave function of the ``string'' in terms of our 
objects(=bits, but in $X_R$ and $X_L$ seperately), we could replace the 
`string'' by thought upon structure provided we take its internal {\bf wave 
function} in terms of the objects to {\bf fall off} exponentially 
(as Gaussians).

The suppression of large transverse momenta in scattering in the faked way 
is rather easily to understand, when one has assumed the wave function 
of the ``string'' in terms of objects as being a Gaussian falling off one.
Then 
the large momentum faked scattering cannot be achieved, because 
it would require some constituents $\sim$ ``objects'' to have large 
transverse momenta, but they do not have that. Or rather it is only 
exponentially few objects with large transverse momentum compared to the 
ones with small  momenta.

\section{Conclusion}
{\bf Conclusion}
We have presented a ``novel 
string field theory'' or perhaps better 
to say ``A solution of several-strings-string-theory'':
\begin{itemize}
\item Our model is string-bit-theory 
deviating in important way from Charles 
Thorn's one.
\item It is a string field theory in the 
sense of describing possibly an arbitrary 
number of strings. But it is very different 
in the spirit of the formalism from usual
string field theories, such as Kaku and Kikkawa
or Wittens string field theories\cite{Kaku3}.

\item Our formulation can be considered a 
{\bf solution} of string theory, in the 
sense that there is no more time development 
left in our object formulation, meaning 
that we solved the equations of motion.
\item All the time development is 
{\bf fake} especially the {\bf scattering}.
\item There is so little ``string'' in 
basic formulation only being a massless 
second quantized particle, that one would 
say: The string is a mathematical construction
from rather trivial starting formalism.
\end{itemize}
{\bf Conclusion on Generalization to other Finite theories (hope)}

Concerning the general looking for finite and therefore meaningfull 
potential theories for all of physics:

\begin{itemize}
\item By analogy with a hadronic bound state, we argued that 
the ultraviolet causing high (transverse) momenta should be avoided by 
making the {\bf Bjorken $x$ for all constituent zero!} (then there would 
even be 
no enrgy for the partons to scatter at all.
\item A model with only zero Bjorken $x=0$ of course would need an 
infinite number of constituents (since the Bjorken x's must add up 
to unity.
\item The string can with our objects
be considered such an 
infinite nummber of constituents model.
\end{itemize}
{\bf Conclusion on Generalization continued.}

\begin{itemize}
\item Could one generalize the string to another pattern
for the 
constituents (than our cyclically ordered sets the doubled strings) ?
\item The p-adic string\cite{padic6, Branko} is the obvious candidate for such  generalization\cite{NN7}. 
\end{itemize}


{\bf Technical Resume Conclusion.}

\begin{itemize}
\item The basic formalism is a massless 
quantum field theory without interactions.
\item These particles are called ``even objects''
and are identified with the even numbered 
constituents/string-bits, when one 
first discretize the string after having 
divided it into right mover and left mover 
d.o.f.   
\end{itemize}


\section*{Acknowledgement}

One of us (H.B.N.) acknowledges the 
Niels Bohr Institute for allowing him to work 
as emeritus and for partial economic support.
Also thanks food etc. support from the Corfu conference
and to Norma Mankoc Borstnik for asking for a 
way to get meaningful quantum field theories in higher than 4 
dimensions. The thinking on hadronic like bound states could namely be looked upon 
as an attempt to find such a scheme using bound states as the theory behind the 
particles for which to make the convergent theory.

M. Ninomiya acknowledges Yukawa Institute of Theoretical Physics, 
Kyoto University, and also   the Niels Bohr Institute
and Niels Bohr International Academy for giving him
 very good hospitality
during his stay. M.N. also acknowledges at 
Yuji Sugawara Lab. 
Science and Engeneering, 
Department of physics sciences Ritsumeikan 
University, 
Kusatsu Campus for allowing him 
 as a visiting 
Researcher.

\end{document}